\documentclass[prb,preprint]{revtex4}

\usepackage{epsfig,wrapfig}

\usepackage{amsmath}

\begin{document}

\title{Student understanding of quantum mechanics at the beginning of graduate instruction}

\author{Chandralekha Singh}

\affiliation{Department of Physics and Astronomy, University of Pittsburgh, Pittsburgh, Pennsylvania 15260}

\begin{abstract}
A survey was developed to probe student understanding of quantum mechanics at the beginning of graduate instruction. 
The survey was administered to 202 physics graduate students enrolled in
first-year quantum mechanics courses from seven universities at the beginning of the first semester.
We also conducted one-on-one interviews with fifteen graduate or advanced undergraduate students who had just completed
a course in which all the content on the survey was covered.
Although students from some universities performed better on average than others,
we found that students share universal difficulties understanding the concepts of quantum mechanics. 
The difficulties were often due to over-generalizations of concepts learned in one context to other contexts where they are
not directly applicable. Difficulties in distinguishing between closely related concepts and making sense of the 
formalism of quantum mechanics were common.
The results of this study can
sensitize instructors of first-year graduate quantum physics to some of the difficulties students are likely to face.
\end{abstract}

\maketitle

\section{Introduction}

A solid understanding of quantum mechanics is essential for most scientists and 
engineers. Single-electron transistors, superconducting quantum interference devices, quantum well lasers, and other 
devices are made possible by the underlying quantum processes.\cite{science}
However, quantum physics is a difficult and abstract subject.\cite{griffiths} 
Unlike classical physics where position and momentum are deterministic variables, in quantum mechanics
they are operators that act on a wavefunction which lies in an abstract Hilbert space. 
In addition, according to the Copenhagen interpretation which is usually taught in quantum courses, an electron in a Hydrogen atom
does not have a definite distance from the nucleus; it is the act of measurement that collapses the wavefunction and 
makes it localized at a certain distance. If the wavefunction is known right before the measurement, 
quantum theory only provides the probability of measuring the distance in a narrow range.

Students taking quantum mechanics often develop survival strategies for performing
reasonably well in their course work. For example, they become proficient at solving the time-independent Schroedinger equation 
with a complicated potential energy and boundary conditions. 
However, students often struggle to make sense of the material and build a robust knowledge structure.
They have difficulty mastering concepts and applying the formalism to answer qualitative questions related to
the general formalism, measurement of physical observables, time-development of the wavefunction, the 
meaning of expectation values, stationary states, and properties of wavefunctions for example.

Several visualization tools have been developed to help students gain intuition about quantum mechanics
concepts.\cite{goldberg,brandt,thaller,hiller,pqm,mario} Recently, low-cost laboratory experiments have been developed
to introduce students to more contemporary quantum ideas.\cite{galvez,havel} Relating
these activities to research on student difficulties can lead to the development of research-based tools that 
can greatly enhance their effectiveness.

Student difficulties in learning physics concepts can be broadly classified in
two categories: gaps in students' knowledge and misconceptions.
Knowledge gaps can be due to a mismatch between the
level at which the material is presented and student's prior knowledge.\cite{simon}
Misconceptions can also impede the learning process at all levels of instruction.\cite{arons,lillian}
Without curricula and pedagogies that appropriately account for common difficulties, instruction is likely to be ineffective.

Several investigations have strived to improve the teaching and learning of quantum 
mechanics.\cite{styer,zollman,redish,robinett,narst,fischler,ireson,niedderer,muller,lawless,today,my,my2,my3}
Styer\cite{styer} has documented several common misconceptions in quantum mechanics. 
Zollman et al.\cite{zollman,narst,vqm} have proposed that quantum concepts be introduced much earlier in physics course sequences and
have designed tutorials and visualization tools which illustrate concepts that can be used at a variety of levels. 
Redish et al.\cite{redish,narst,vqm} have conducted investigations of student difficulties and
developed research-based material to teach quantum physics concepts to a wide range of science and engineering students.
Robinett et al.\cite{robinett} designed a test related to quantum physics concepts that can be 
administered to students in courses ranging from introductory quantum physics to graduate quantum mechanics. 
Several other investigations have focused on students' conceptions about modern physics early in college or at the
pre-college level.\cite{narst,fischler,ireson,niedderer,muller,lawless}

One of our earlier investigations explored student understanding of 
quantum measurement and time-dependence of the expectation value.\cite{today,my,my2,my3}
Our analysis of the data obtained from the written test and interviews
showed that advanced students have common difficulties and misconceptions 
independent of their background, teaching style, textbook, and institution, analogous to the patterns of misconceptions 
observed in introductory physics courses. 

\section{Methodology}

Here we describe the findings from a graduate quantum mechanics survey which was developed and administered to 202 graduate
students from seven universities in the United States. 
The 50 minute written survey administered at the beginning of a first-year, first-semester/quarter 
graduate quantum mechanics course covers a range of concepts. 
To understand the reasoning difficulties in depth, we also interviewed fifteen graduate or
advanced undergraduate students enrolled in a quantum mechanics course at the university of Pittsburgh 
in which all of the concepts on the survey were covered. 
Although students from some universities performed better on average than others, we find that students have common conceptual
difficulties regardless of where they are enrolled.
By conceptual difficulties, we refer to difficulties in using
one's knowledge to interpret, explain, and draw inferences while answering
qualitative questions in different contexts. 

During the design of the survey we consulted three Pittsburgh faculty members
who had taught quantum mechanics. Previously, we discussed with them the concepts that they expected the
students in a first-semester graduate quantum mechanics to know. Because the first year graduate students at Pittsburgh are given a placement test 
in each of the core courses to determine whether they are better suited for the graduate or the corresponding undergraduate course,
we discussed the kinds of questions instructors would put on the placement test. 
The initial longer version of the survey was designed based on the
concepts the instructors considered important prior knowledge for the graduate course.
In addition to commenting on the wording of the questions to eliminate ambiguity, we asked the faculty members to rate the 
questions and comment on what should 
be included in the survey. Based upon their feedback, we iterated the survey questions several times
before using a version that was individually administered to a few graduate students. We discussed the survey with the students
after they had taken it and fine-tuned it based on their responses. 
As shown in the Appendix, the graduate survey covers topics related to
the time-dependent Schroedinger equation and time-independent Schroedinger equation, the time dependence of the wavefunction, the
probability of measuring energy and position, the expectation values of the energy, the identification of
allowed wavefunctions for an infinite square well (not the stationary states), 
graphical representations of the bound and scattering states for a finite square well, and the
formalism associated with the Stern-Gerlach experiment.

The universities that participated in this study and the number of students from each university are given in Table~\ref{1}.
For four of the seven universities, the survey was administered in the first class period in the first week ,
while in the other three universities, it was administered in the second or third weeks.
At some universities, students were either given the survey as a placement test, or given a small number of bonus points for taking the survey regardless of their actual 
performance on the survey, or were asked to take the survey seriously because the results could help 
tailor the graduate instruction and help in making the undergraduate quantum mechanics courses more effective.
At universities where the test was not given as a placement test, the students were told ahead of time
that they will be taking a survey on the material covered in the undergraduate 
quantum mechanics in case they wanted to review the material.

To investigate the difficulties with these concepts in more depth,
fifteen students (graduate students or physics seniors from the University of Pittsburgh 
in an undergraduate quantum mechanics course in which all of the survey content was covered) were interviewed
using a think-aloud protocol.\cite{chi} These interviews were semi-structured in the sense that we had a list of issues related to each
question that we definitely wanted to probe. These issues were not brought up initially because we wanted to give students an 
opportunity to formulate their own response and reasoning. The students were asked to verbalize their thoughts while they were working on the survey questions. 
They were not interrupted unless they remained quiet for a while in which case they were reminded to ``keep talking.'' 
After students had finished articulating their ideas, 
they were asked further questions to clarify issues. Some of this later probing was from the list of issues that we had planned to probe
initially (and asked students at the end if they did not bring it up themselves) and others were questions designed on-the-spot
to get a better understanding of a particular student's reasoning.

\section{Discussion}

The Appendix shows the final version of the graduate survey. 
Table~\ref{2} shows the responses in percentage of students for each question. 
These difficulties are categorized and discussed in detail in the following.

\subsection{Time-independent Schroedinger equation is most fundamental}

One difficulty that was pervasive across several questions was the overemphasis on the time-independent Schroedinger equation.
For example, in Question~1 students were asked to write down the most fundamental equation of quantum mechanics. 
We were expecting that students would write down some form of the time-dependent Schroedinger equation
\begin{subequations}
\begin{align}
i \hbar \frac{\partial \vert \Psi(t) \rangle}{\partial t} & =\hat H \vert \Psi(t) \rangle,\\
\noalign{\noindent or}
i \hbar \frac{\partial \Psi(x,t)}{\partial t}& =\hat H \Psi(x,t). \label{1b}
\end{align}
\end{subequations}
Equation~\eqref{1b} is the time-dependent Schroedinger equation in one spatial dimension for which the Hamiltonian 
$\hat H=\hat p^2/(2m)+V(\hat x)$.
Responses that cited the position-momentum uncertainty principle 
(3 students) or the commutation relation between position and momentum (1 student) as the most fundamental equation of quantum mechanics
were also considered correct. 
Even if students did not explicitly write down the Hamiltonian in terms of the potential and kinetic energy operators, 
their responses were
considered correct. We also considered the response correct if the students made mistakes such as forgetting $\hbar$, the relative signs of 
various terms in the equation, and the mass of the particle $m$ in the Hamiltonian. 
Only $32\%$ of the students provided a correct response with this scoring criterion. 

Table~\ref{2} shows that $48\%$ of the students believed 
that the time-independent Schroedinger equation $H \phi_n=E_n \phi_n$ is the most fundamental equation of quantum mechanics 
(not all of the equations given by students had the Hamiltonian written correctly). 
It is correct that if the potential energy is time-independent, we can use separation of variables 
to obtain the time-independent Schroedinger equation which is an eigenvalue equation for the Hamiltonian. The eigenstates of $\hat H$ obtained by solving the time-independent Schroedinger equation are
the stationary states which form a complete set of states. Most advanced undergraduate
quantum mechanics courses de-emphasize the time-dependent Schroedinger equation,
and students recall the quantum mechanics course as an exercise in solving time-independent Schroedinger equation.
As we will discuss, an overemphasis on time-independent Schroedinger equation also leads to the main difficulty with quantum dynamics.

In Question~4 students were asked to explain why they agree or disagree with the following statement: ``By definition, the 
Hamiltonian acting on {\it any} state of the system $\psi$ will give the same state back, that is, $\hat H \psi=E \psi$." We wanted students
to disagree with the statement and note that it is only true if $\psi$ is a stationary state.
In general, $\psi=\sum_{n=1}^{\infty} C_n \phi_n$, where $\phi_n$ are the stationary
states and $C_n=\langle \phi_n \vert \psi \rangle$. Then, $\hat H \psi=\sum_{n=1}^{\infty} C_n E_n \phi_n \ne E \psi$. Just writing down ``disagree" was not enough for the response to be counted correct. Students had to provide the correct reasoning.
Only $29\%$ of the students provided the correct response. 
Thirty-nine percent of students wrote incorrectly that the statement is unconditionally
correct. This percentage is
slightly lower than the percentage of students who claimed in Question~1 that $\hat H \psi=E \psi$ is the most fundamental equation of 
quantum mechanics. 
Typically, these students were confident of their responses as can be seen from these examples: (a) ``Agree. This is what 80 years of experiment has proven. If future experiments prove this statement wrong,
then I'll update my opinion on this subject.'' (b)
``Agree, this is a fundamental postulate of quantum mechanics which is proved to be highly exact until present.'' (c) ``Agree. This is what Schroedinger equation implies and it is what quantum mechanics is founded on.''
(d) ``Agree. $\hat H$ commutes with all operators which measure observable quantities.
Hence, any state $\Psi$ is an eigenstate of the system.''

During the interview, one student said ``It seems like you are asking this question because you want me to disagree with this statement.
Unfortunately, it seems correct to me." After reading Question~4, another interviewed student who had earlier 
said that the Schrodinger equation is the most fundamental
equation of quantum mechanics, but could not remember the equation said ``Oh, here is the answer to Question~1.''
Our earlier investigation\cite{today,my,my2,my3} found that the difficulties related to the time dependence of expectation values are also
often related to applying the properties of stationary states to non-stationary states (such as the eigenstates of position or 
momentum operators).

\subsection{Hamiltonian acting on a state represents energy measurement}

Eleven percent of the students answering Question~4 believed incorrectly that any statement involving a 
Hamiltonian operator acting on a state is a statement about the measurement of energy.
Some of these students who incorrectly claimed that $\hat H \psi=E \psi$ is a statement about energy measurement agreed
with the statement while others disagreed.
Those who disagreed often claimed that $\hat H \psi=E_n \phi_n$, 
because as soon as $\hat H$ acts on $\psi$, the wavefunction will
collapse into one of the stationary states $\phi_n$ and the corresponding energy $E_n$ will be obtained. 
The following examples are typical of students with this misconception: (a) ``Agree. $\hat H$ is the operator for an energy measurement.
Once this measurement takes place, the specific value $E$ of the energy will be known.'' (b) ``Agree. If you make a measurement of energy by applying $H$ to a state of an electron in hydrogen atom you will get the energy.'' (c) ``Agree except when the system is in a linear superposition. In that case, Hamiltonian acting on it will make it settle
into only one of the term corresponding to the measured energy.'' (d) ``Hamiltonian acting on a system will collapse the system into one of the possible energy states. This does not give the
same original state unless the previous state was same as resulting state.'' (e) ``Disagree. The Hamiltonian acting on a mixed state will single out one component.
The wavefunction will collapse to a different state once the energy has been determined to be that of one component.''

The interviews and written answers suggest that these students believed that 
the measurement of a physical observable in a particular state 
is achieved by acting with the corresponding operator on the state.
The incorrect notions are overgeneralizations of the fact that
after the measurement of energy, the system is in a stationary state so $\hat H \phi_n=E_n \phi_n$.
This example illustrates the difficulty students have in relating the formalism of quantum mechanics to the 
measurement of a physical observable. 

\subsection{$\hat Q \psi=\lambda \psi$ for any physical observable $Q$}

Individual interviews related to Question~4 suggest that some students believed that 
if an operator $\hat Q$ corresponding to a physical observable $Q$ acts on any state $\psi$, it will
yield the corresponding eigenvalue $\lambda$ and the same state back, that is, $\hat Q \psi=\lambda \psi$.
Some of these students were overgeneralizing their ``$\hat H \psi=E \psi$" reasoning and attributing
$\hat Q \psi=\lambda \psi$ to the measurement of an observable $Q$.
Before overgeneralizing to any physical observables, these students often agreed with the $\hat H \psi=E \psi$
statement with arguments such as ``the Hamiltonian is the quantum mechanical operator which corresponds to the 
physical observable energy" or ``if $H$ did not give back the same state it would not be a hermitian operator 
and therefore would not correspond to an observable." Of course, $\hat Q \psi \ne \lambda \psi$ unless $\psi$
is an eigenstate of $\hat Q$ and in general
$\psi=\sum_{n=1}^{\infty} D_n \psi_n$, where $\psi_n$ are the eigenstates of $\hat Q$
and $D_n=\langle \psi_n \vert \psi \rangle$. Then, $\hat Q \psi=\sum_{n=1}^{\infty} D_n \lambda_n \psi_n$ (for an
observable with a discrete eigenvalue spectrum).

\subsection{$\hat H \psi=E \psi$ if $\hat H$ does not depend on time}

In response to Question~4, $10\%$ of the students agreed with the statement as long as 
the Hamiltonian is not time-dependent.
They often claimed incorrectly that if $\hat H$ is not time-dependent, 
the energy for the system is conserved so $\hat H \psi=E \psi$ must be correct. 
The following are typical examples: (a) ``Agree, if the potential energy does not depend on time.'' (b) ``Agree but only if the energy is conserved for this system.''
(c) ``Agree because energy is a constant of motion.''
(d) ``Agree if it is a closed system because $H$ is a linear operator and gives the same state back multiplied by the energy.''

Although the energy is conserved if the Hamiltonian is time-independent, $\hat H \psi=E \psi$ need not be true. For example,
if the system is in a linear superposition of stationary states, $\hat H \psi\ne E \psi$ although the energy is conserved.

\subsection{Difficulties related to the time-development of wavefunction}

The most common difficulties with quantum dynamics are coupled with an overemphasis on the time-independent Schroedinger equation.
Equation~(1) shows that the evolution of the wavefunction $\Psi(x,t)$ is governed by the Hamiltonian $\hat H$ of the system via the
time-dependent Schroedinger equation and there is no dynamics in the time-independent Schroedinger equation.
Question~2 concerned an electron in a one-dimensional infinite square well, initially ($t=0$) in a linear
superposition of the ground state $\phi_1(x)$ and the first excited state $\phi_2(x)$.
In Question~2(a), students were asked to write down the wavefunction $\Phi(x,t)$ at time $t$. We were expecting the following response:
$\Psi(x,t)=\sqrt{2/7} \phi_1(x) e^{-iE_1t/\hbar}+\sqrt{5/7} \phi_2(x) e^{-iE_2t/\hbar}$. 
Responses were considered correct if students wrote the phase factor 
for the first term as $e^{-iAE_1t}$, where $A$ is any real constant (for example, $\hbar$ in the numerator, incorrect sign, or some other constant, 
for example, mass $m$ in the phase were considered minor problems and ignored even though they can make the phase a quantity with dimension). 
Some students wrote incorrect intermediate steps; for example,
$\Psi(x,t)=\Psi(x,0)e^{-iEt/\hbar}=\sqrt{2/7} \phi_1(x) e^{-iE_1t/\hbar}+\sqrt{5/7} \phi_2(x) e^{-iE_2t/\hbar}$. 
Such responses were considered correct.
During the individual interviews, a student proceeded from an intermediate incorrect step to the correct time-dependence in the second step
similar to the above expression. Further probing showed that the student was having difficulty distinguishing between the Hamiltonian
operator and its eigenvalue and was probably thinking of
$\Psi(x,t)=e^{-i\hat H t/\hbar}\Psi(x,0)=\sqrt{2/7} \phi_1(x) e^{-iE_1t/\hbar}+\sqrt{5/7} \phi_2(x) e^{-iE_2t/\hbar}$, where
the Hamiltonian $\hat H$ acting on the stationary states gives the corresponding energies.

As shown in Table~\ref{1}, $31\%$ of students wrote common phase factors for both terms, for example,
$\Psi(x,t)=\Psi(x,0)e^{-iEt/\hbar}$. Interviews suggest that these students were having difficulty distinguishing
between the time-dependence of stationary and non-stationary states. Because the Hamiltonian operator governs the time-development
of the system, the time-dependence of a stationary state is via a simple phase factor. In general non-stationary states 
have a non-trivial time-dependence because each term in a linear superposition of stationary states evolves via a different phase factor.
Apart from using $e^{-iEt/\hbar}$ as the common phase factor, other common choices include $e^{-i \omega t}$, $e^{-i \hbar t}$,
$e^{-it}$, $e^{-ixt}$, and $e^{-ikt}$.

Interestingly, $9\%$ of the students believed that $\Psi(x,t)$ should not have any time dependence; during the interviews
some students justified their claim by pointing to the time-independent Schroedinger equation and adding that the Hamiltonian is not time-dependent.
Several students thought that the time dependence was a decaying exponential,
for example, of the type $\Psi(x,0)e^{-xt}$, $\Psi(x,0)e^{-Et}$, $\Psi(x,0)e^{-ct}$, or $\Psi(x,0)e^{-t}$. During the interviews
some of these students explained their choice by insisting that the wavefunction must decay with time because
``this is what happens for all physical systems."
Other incorrect responses were due to the partial retrieval of related facts from memory such (a)
$\sqrt{2/7} \phi_1(x+\omega t)+\sqrt{5/7} \phi_2(x+\omega t)$, (b)
$\sqrt{2/7} \phi_1(x) e^{-i\phi_1 t}+\sqrt{5/7} \phi_2(x) e^{-i\phi_2 t}$, (c)
$\sqrt{2/7} \phi_1(x) e^{-ix t}+\sqrt{5/7} \phi_2(x) e^{-i 2 x t}$, and (d)
$\sqrt{2/7} \phi_1(x) \sin (\pi t)+\sqrt{5/7} \phi_2(x) cos(2 \pi t)$.
The interviews suggest that these students often correctly remembered that the time-dependence of non-stationary states cannot be
represented by a common time-dependent phase factor, but did not know how to correctly evaluate $\Psi(x,t)$.

\subsection{Difficulties with measurement and expectation value}

\subsubsection{Difficulty interpreting the meaning of expectation value}

Although Question~2(b) was the easiest on the survey with $67\%$ correct
responses, a comparison with the response for Question~2(c), for which $39\%$ provided the correct response,
is revealing. It shows that many students who can calculate probabilities for the
possible outcomes of energy measurement were unable to use that information
to determine the expectation value of the energy. 
In Question~2(b) students were asked about the possible values of the energy of the electron
and the probability of measuring each in an initial state $\psi(x,0)$. 
We expected students to note that the only possible values of the energy in state 
$\psi(x,0)$ are $E_1$ and $E_2$ and their respective probabilities are $2/7$ and $5/7$.
In Question~2(c), students had to calculate the expectation value of the energy in the state 
$\Psi(x,t)$. The expectation value of the energy is
time-independent 
because the Hamiltonian does not depend on time. If $\Psi(x,t)=C_1(t) \phi_1(x)+C_2(t) \phi_2(x)$, 
then the expectation value of the energy in this state is 
$\langle E \rangle=P_1 E_1+P_2 E_2=\vert C_1(t)\vert^2 E_1+ \vert C_2(t)\vert^2 E_2=(2/7)E_1 +(5/7)E_2$,
where $P_i=\vert C_i(t)\vert^2$ is the probability of measuring the energy $E_i$ at time $t$. 

Many students who answered Question~2(b) correctly
(including those who also answered Question~(c) correctly) calculated $\langle
E\rangle $ by brute-force: first writing $\langle E\rangle
=\!\int_{-\infty }^{+\infty }\Psi^{\ast}\hat{H}\Psi dx$, expressing $\Psi(x,t)$ 
in terms of the linear superposition of two energy eigenstates, then acting $\hat{H}$ on
the eigenstates, and finally using orthogonality to obtain the answer.
Some got lost early in this process, and others did not remember some other
mechanical step, for example, taking the complex conjugate of the wavefunction,
using the orthogonality of stationary states, or not realizing the proper limits
of the integral.

The interviews reveal that many did not know or recall the interpretation of
expectation value as an ensemble average and did not realize that
expectation values could be calculated more simply in this case by taking
advantage of their answer to Question~2(b). Some believed that the expectation value of the
energy should depend on time (even those who correctly evaluated $\Psi (x,t)$
in Question~2(a)).

In the interview one student who answered Question~2(b) correctly, did not know how to apply it to Question~2(c). He wrote an explicit
expression involving the wavefunction for the ground and first excited states, but thought that $H\phi_n=E_n$ with no
$\phi_n$ on the right-hand side of this equation. Therefore, he found a final expression for $\langle E \rangle$ that involved wavefunctions. When he was
told explicitly by the interviewer that the final answer should not be in terms of $\phi_1$ and $\phi_2$ and he should try to find his
mistake, the student could not find his mistake. The interviewer then explicitly pointed to the particular step in which
he had made the mistake and asked him to find it. The student still had difficulty because he believed $H\phi_n=E_n$ was correct. 
Finally, the interviewer told the student that
$H\phi_n=E_n \phi_n$. At this point, the student was able to use orthonormality correctly to obtain the correct result
$\langle E \rangle=(2/7)E_1 +(5/7)E_2$. Then, the interviewer asked him to think about whether it is possible to calculate
$\langle E \rangle$ based on his response to Question~2(b). The student's eyes brightened and he responded, ``Oh yes \ldots I never thought of
it this way\ldots I can just multiply the probability of measuring a particular energy with that
energy and add them up to get the expectation value because expectation value is the average value.'' 
Then, pointing to his detailed work for Question~2(c) he added, ``You can see that the time dependence cancels out \ldots.''
$17\%$ of the students simply wrote $\langle \Psi \vert E \vert \Psi \rangle$ or $\langle \Psi \vert H \vert \Psi \rangle$ and did not 
know how to proceed. A few students wrote the expectation value of energy as $[E_1+E_2]/2$ or $[(2/7)E_1 +(5/7)E_2]/2$.

\subsubsection{Difficulty interpreting the measurement postulate}

According to the Copenhagen interpretation, the measurement of a physical observable 
instantaneously collapses the state to an eigenstate of the corresponding operator.
In Question~2(d), students were asked to write the wavefunction right after the measurement of energy if the measurement
yields $4\pi^2 \hbar^2/(2ma^2)$. Because the measurement yields the energy of the first excited state, the measurement collapsed the
system to the first excited state $\phi_2(x)$. 
Students did well on this question with $79\%$ providing the correct response. However, the $79\%$ includes those students ($17\%$)
who wrote the unnormalized wavefunction $\sqrt{5/7} \phi_2(x)$. Interviews suggest that some students who did not
normalize the wavefunction strongly believed that the coefficient $\sqrt{5/7}$ must be included explicitly to represent the
wavefunction after the measurement
not realizing that state $\phi_2(x)$ is already normalized. 

In response to Question~2(d), 
some students thought that the system should remain in the original state, which is a linear superposition of the 
ground and first excited states. 
One of the interviewed students said ``Well, the answer to this question depends upon how much time you wait after the measurement. If you are talking about what happens at the instant you measure the energy,
the wavefunction will be $\phi_2$ but if you wait long enough it will go back to the state before the measurement.'' The notion that the
system must go back to the original state before the measurement was deep-rooted in the student's mind and could not be dislodged
even after the interviewer asked several further questions about it. When the interviewer said that it was not clear why that would
be the case, the student said, ``The collapse of the wavefunction is temporary \ldots\ Something has to happen to the wavefunction for
you to be able to measure energy or position, but after the measurement the wavefunction must go back to what it {\it actually} 
(student's emphasis) is supposed to be.'' When probed further, the student continued, ``I remember that if you measure 
position you will get a delta function, but it will stay that way only if you do repeated measurement \ldots\ if you let it 
evolve it will go back to the previous state (before the measurement)."

Some students confused the measurement of energy with the measurement of position and
drew a delta function in Question~2(e). They 
claimed that the wavefunction will become very peaked about a given position after the energy measurement. 
An interviewed student drew a wavefunction which was a delta function in position. 
He claimed incorrectly that because the energy will have a definite value after the measurement
of energy, the wavefunction should be localized in position.
Further probing showed that he was confused about the vertical axis of his plot and said that he may be plotting energy along
that axis. When asked
explicitly about what it means for the energy to be localized at a fixed position, he said he may be doing something wrong but he
was not sure what else to do.

\subsubsection{Confusion between the probability of measuring position and the expectation value of position}

Born's probabilistic interpretation of the wavefunction can also be confusing for students.
In Question~2(f), students were told that immediately after the measurement of energy, a measurement of the electron position is performed. 
They were asked to
describe qualitatively the possible values of the position they could measure and the probability of measuring them. We hoped that students would
note that it is possible to measure position values between $x=0$ and $x=a$ (except at $x=0$, $a/2$, and $a$ where the wavefunction is zero), and
according to Born's interpretation, $\vert \phi_2(x) \vert^2 dx$ gives the probability of finding the particle e between $x$ and $x+dx$. Only $38\%$ of the students
provided the correct response. Partial responses were considered correct for tallying purposes
if students wrote anything that was correctly related to the above wavefunction, for example, 
``The probability of finding the electron is highest at $a/4$ and $3a/4$," or ``The probability of finding the electron is 
non-zero only in the well.''

Eleven percent of the students tried to find the expectation value of position $\langle x \rangle$ instead of the probability of finding the electron
at a given position. They wrote the expectation value of position in terms of an integral involving the wavefunction. Many of them
explicitly wrote that $\mbox{Probability}=\!(2/a)\!\int_0^a x \sin^2(2\pi x/a) dx$ and
believed that instead of $\langle x \rangle$ they were calculating the probability of measuring the position of electron.
During the interview, one student said (and wrote) that the probability is
$\int\!x\,\vert \Psi \vert^2 dx$. When the interviewer asked why $\vert \Psi \vert^2 $ should be multiplied by
$x$ and if there is any significance of $\vert \Psi \vert^2 dx$ alone, the student said, 
``$\vert \Psi \vert^2$
gives the probability of the wavefunction being at a given position and if you multiply it by $x$ you get the probability of 
{\it measuring} (student's emphasis) the position $x$.'' When the student was asked questions about the meaning of the 
``wavefunction being at a given position," and the purpose of the integral and its
limits, the student was unsure. He said that the reason he wrote the integral is because
$x\,\vert \Psi \vert^2 dx$ without an integral looked strange to him. 

\subsubsection{Other difficulties with measurement and expectation value}

Other difficulties with measurements were observed as well. For example,
in response to Question~2(f), 7\% tried to use the (generalized) uncertainty principle between energy and position or between position and momentum,
but most of their arguments led to incorrect inferences. For example, several students noted that because the energy is well-defined immediately after the measurement
of energy, the uncertainty in position must be infinite according to the uncertainty principle. Some students even went on to argue that the probability
of measuring the particle's position is the same everywhere. Others restricted themselves only to the inside of the well and noted that the uncertainty
principle says that the probability of finding the particle is the same everywhere inside the well and for each value of position inside the well this
constant probability is $1/a$.
For example, one student said, ``Must be between $x=0$ and $x=a$ \ldots but by knowing the exact energy, we can know nothing about position so probable
position is spread evenly across in $0<x<a$ region.''
Some students thought that the most probable values of position were the only possible values of the position that can be measured.
For example, one student said ``According to the graph above (in Question~2(e)), we can get positions $a/4$ and $3a/4$ each with individual probability $1/2$.''
The following statement was made by a student who believed that it may not be possible to measure the position after measuring the energy: 
``Can you even do that? Doesn't making a measurement change the system in a manner that makes another measurement invalid?"
The fact that the student believed that making a measurement of one observable 
can make the immediate measurement of another
observable invalid, sheds light on student's epistemology about quantum theory.

Seven percent of students answering Question~2(b) became confused between individual measurements of the energy and its expectation value, and
almost none of these students calculated the correct expectation value of the energy. 
Another common mistake was assuming that all allowed energies
for the infinite square well were possible and the ground state is the most probable because it is the lowest energy state.
Some students thought that the probabilities for measuring $E_1$ and $E_2$ are $4/(7a)$ and $10/(7a)$ respectively 
because they included the normalization factor for the stationary state wavefunctions $\sqrt{2/a}$ while squaring the coefficients.
Some thought that the probability amplitudes were the probabilities of measuring energy and did not square the coefficients
$\sqrt{2/7}$ and $\sqrt{5/7}$.

\subsection{Difficulty in determining possible wavefunction}

Any smooth function that satisfies the boundary condition for a system is a possible wavefunction.
We note that question (3) asks if the three wavefunctions given are allowed for an electron in a one-dimensional 
infinite square well, interviews suggest that students correctly interpreted it to mean whether they were
possible wavefunctions.
In this Question, we hoped that students would note that the wavefunction $A e^{-{((x-a/2)/a)}^2}$ is not possible because 
it does not satisfy the boundary
conditions (does not go to zero at $x=0$ and $x=a$). The first two wavefunctions
$A\sin^3(\pi x/a)$, $A[\sqrt{2/5} \sin(\pi x/a)+ \sqrt{3/5} \sin(2\pi x/a)]$ with suitable normalization constants are both smooth functions that satisfy the boundary condition (each of them goes to zero at $x=0$ and $x=a$) so each
can be written as a linear superposition of the two stationary states. Seventy-nine percent of the students could identify that the second wavefunction is a possible wavefunction because it is explicitly written in the form of a linear superposition of stationary states.
Only $34\%$ gave the correct answer for all three wavefunctions. Within this subset,
a majority correctly explained their reasoning based on whether the boundary conditions are satisfied by these wavefunctions.
For tallying purposes, responses were considered correct even if the reasoning was not completely correct.
For example, one student wrote incorrectly: ``The first two wavefunctions are allowed because they satisfy the equation
$\hat H \Psi=E \Psi$ and the boundary condition works.'' The first part of the reasoning provided by this student is incorrect while the
second part that relates to the boundary condition is correct.

Forty-five percent believed that $A\sin^3(\pi x/a)$ is
not possible but that $A[\sqrt{2/5} \sin(\pi x/a)+ \sqrt{3/5} \sin(2\pi x/a)]$ is possible.
The interviews suggest that a majority of students did not know that any smooth single-valued wavefunction that satisfies the boundary
conditions can be written as a linear superposition of stationary states.
Interviews and written explanations suggest that many students incorrectly believed that the following two constraints
must be independently satisfied for a wavefunction to be a possible wavefunction: it must be a smooth single-valued function that satisfies the boundary conditions and it must either be possible to write it as a linear superposition of stationary states, or it must satisfy the time-independent
Schroedinger equation.

As in the following example, some who correctly realized that $A\sin^3(\pi x/a)$ satisfies the boundary condition,
incorrectly claimed that it is still not a possible wavefunction: ``$A\sin^3(\pi x/a)$ satisfies b.c. but does not satisfy Schrodinger equation (that is, it cannot represent a particle wave).
The second one is a solution to S.E. (it is a particle wave). The third does not satisfy b.c.''

Many claimed that only pure sinusoidal wavefunctions are possible, and $\sin^2$ or $\sin^3$ are not possible.
The interviews and written explanations suggest that students believed that $A\sin^3(\pi x/a)$ cannot be written as a
linear superposition of stationary states and hence it is not a possible wavefunction. The following are examples: (a) ``$A\sin^3(\pi x/a)$ is not allowed because it is not an eigenfunction nor a linear combination.'' (b) ``$A\sin^3(\pi x/a)$ is not allowed because it is not a linear function but Schroedinger equation is linear.'' (c) ``$A\sin^3(\pi x/a)$ is not allowed. Only simple sines or cosines are allowed.'' (d) ``$A\sin^3(\pi x/a)$ works for 3 electrons but not one.''

The most common incorrect response claimed incorrectly that $A\sin^3(\pi x/a)$ is not a possible wavefunction because it
does not satisfy $\hat H\Psi=E \Psi$.
Students asserted that $A\sin^3(\pi x/a)$ does not satisfy the time-independent Schroedinger equation (which they believed was the equation that all possible wavefunctions should
satisfy) but $A[\sqrt{2/5} \sin(\pi x/a)+ \sqrt{3/5} \sin(2\pi x/a)]$ does.
Many explicitly wrote the Hamiltonian as $\frac{-\hbar^2}{2m} \frac{\partial^2}{\partial x^2}$ and showed that the second derivative of
 $A\sin^3(\pi x/a)$ will not yield the same wavefunction back multiplied by a constant.
Incidentally, the same students did not attempt to take the second derivative of $A[\sqrt{2/5} \sin(\pi x/a)+ \sqrt{3/5} \sin(2\pi x/a)]$; otherwise
they would have realized that even this wavefunction does not give back the same wavefunction multiplied by a constant.
For this latter wavefunction, a majority claimed that it is possible
because it is a linear superposition of $\sin(n\pi x/a)$.
Incidentally, $A\sin^3(\pi x/a)$ can also be written as a linear superposition of
only two stationary states.
Thus, students used different reasoning to test the validity of the first two wavefunctions as in the following example: $\frac{-\hbar^2}{2m} \frac{\partial^2}{\partial x^2}A\sin^3(\pi x/a)$ cannot be equal to $E\,A\sin^3(\pi x/a)$ so it isn't acceptable.
Second is acceptable because it is linear combination of sine.

Some students incorrectly noted that $A[\sqrt{2/5} \sin(\pi x/a)+ \sqrt{3/5} \sin(2\pi x/a)]$ is possible inside the well
and $A e^{-{((x-a/2)/a)}^2}$ is possible outside the well.
Others incorrectly claimed that $A\sin^3(\pi x/a)$ does not satisfy the boundary condition for the system
but $A[\sqrt{2/5} \sin(\pi x/a)+ \sqrt{3/5} \sin(2\pi x/a)]$ does.
Some dismissed $A\sin^3(\pi x/a)$ claiming it is an odd function that cannot be a possible wavefunction for an infinite square well which is an even potential.
In the interview, a student who thought that only $A[\sqrt{2/5} \sin(\pi x/a)+ \sqrt{3/5} \sin(2\pi x/a)]$ is possible said, ``these other
two are not linear superpositions.'' When the interviewer asked explicitly how he could tell that the other two wavefunctions cannot be written
as a linear superposition, he said, ``$A\sin^3(\pi x/a)$ is clearly multiplicative not additive \ldots you cannot make a cubic function out of
linear superposition \ldots this exponential cannot be a linear superposition either.''

Five percent of students claimed that $A e^{-{((x-a/2)/a)}^2}$ is
a possible wavefunction for an infinite square well. These students did not examine the boundary condition. They sometimes
claimed that an exponential can be represented by sines and cosines, and hence it is possible or
focused only on the normalization of the wavefunction.
Not all students who correctly wrote that
$A e^{-{((x-a/2)/a)}^2}$ is not a possible wavefunction provided the correct reasoning.
Many students claimed that the possible wavefunctions
for an infinite square well can only be of the form $A \sin(n\pi x/a)$ or that $A e^{-{((x-a/2)/a)}^2}$ is possible
only for a simple harmonic oscillator or a free particle.

\subsection{Difficulty distinguishing between three dimensional space and Hilbert space}

In quantum theory, it is necessary to interpret the outcome of real experiments performed in real space by
making connection with an abstract Hilbert space (state space) in which the wavefunction lies. The
physical observables that are measured in the laboratory correspond to Hermitian operators in the Hilbert space whose
eigenstates span the space. Knowing the initial wavefunction and the Hamiltonian of the system allows
one to determine the time-evolution of the wavefunction unambiguously and the measurement postulate can be used to 
determine the possible outcomes of individual measurements and ensemble averages (expectation values).

It is difficult for students to distinguish between vectors in real space and Hilbert space. For example, 
$S_x$, $S_y$ and $S_z$ denote the orthogonal components of the spin angular momentum vector of an electron in three dimensions, each 
of which is a physical observable that can be measured in the laboratory. However, the Hilbert space corresponding to the
spin degree of freedom for a spin-1/2 particle is two-dimensional (2D). In this Hilbert space, $\hat S_x$, $\hat S_y$ and $\hat S_z$
are operators whose eigenstates span 2D space. 
The eigenstates of $\hat S_x$ are vectors which span the 2D space and are orthogonal to each other
(but not orthogonal to the eigenstate of $\hat S_y$ or $\hat S_z$). 
Also, $\hat S_x$, $\hat S_y$ and $\hat S_z$ are operators and not orthogonal components of a vector in 2D space. If the electron is in a magnetic field with the gradient
in the $z$ direction in the laboratory (real space) as in a Stern-Gerlach experiment, the magnetic field is a vector
in three-dimensional (3D) space but not in 2D space. It does not make sense to compare vectors in 3D space with
vectors in the 2D space as in statements such as ``the magnetic field gradient is perpendicular to the eigenstates of $\hat S_x$.''
These distinctions are difficult for students to make and such difficulties are common as discussed in the following.

Question~5 has two parts, both of which are related to the Stern-Gerlach experiment. 
The notation $\vert \uparrow_z \rangle$ and $\vert \downarrow_z \rangle$ represent the orthonormal eigenstates of 
$\hat S_z$ (the $z$ component of the spin angular momentum) of a spin-1/2 particle.
In one version of this question, a beam of neutral silver atoms with spin-1/2 was sent through the Stern-Gerlach apparatus.
Students had similar difficulties with both versions.
In Question~5(a) a beam of electrons propagating along the $y$ direction (into the page) in spin state $(\vert \uparrow_z \rangle+
\vert \downarrow_z \rangle)/\sqrt {2}$ is sent through a Stern-Gerlach apparatus with a vertical magnetic field gradient in the $-z$ direction. Students were 
asked to sketch the electron cloud pattern that they expected to see on a distant phosphor screen in the $x$-$z$ plane and explain 
their reasoning. We wanted students to realize that the magnetic field gradient in the $-z$ direction would exert a force 
on the electron due to its spin angular momentum and two spots would be observed on the phosphor screen due to the splitting 
of the beam along the $z$ direction corresponding to electron spin component in the
$\vert \uparrow_z \rangle$ and $\vert \downarrow_z \rangle$ states. 
All responses in which students noted that there will be a splitting along the $z$ direction
were considered correct even if they did not explain their reasoning. 
Only $41\%$ of the students provided the correct response. 
Many students thought that there will only be a single spot on the phospor screen as in these typical responses: (a) SGA (the Stern-Gerlag apparatus) will pick up the electrons with spin down because the gradient is in the $-z$ direction.
The screen will show electron cloud only in $-z$ part; (b)
All of the electrons that come out of the SGA will be spin down with expectation value
$-\hbar/2$ because the field gradient is in $-z$ direction; (c) Magnetic field is going to align the spin in that direction so most of the electrons will align along $-z$ direction. We may
still have a few in the $+z$ direction but the probability will be very small.

Students were often confused in the interviews about the origin of the force on the particles and
whether there should be a force on the particles at all as they pass through the Stern-Gerlag apparatus.

In Question~5(b)
a beam of electrons propagating along the $y$ direction (into the page) in spin state 
$\vert \uparrow_z \rangle$ is sent through an apparatus with a horizontal magnetic field gradient in the $-x$ direction. 
Students were asked to sketch the electron cloud pattern they expect to see on a distant phosphor screen 
in the $x$-$z$ plane and explain their reasoning.
This question is more challenging than Question~5(a) because students have to realize that the 
eigenstate of $\hat S_z$, $\vert \uparrow_z \rangle$ can be 
written as a linear superposition of the eigenstates of $\hat S_x$, that is, 
$\vert \uparrow_z \rangle=(\vert \uparrow_x \rangle + \vert \downarrow_x \rangle) /\sqrt{2}$. 
Therefore, the magnetic field gradient in the $-x$ direction will 
split the beam along the $x$ direction corresponding to the electron spin components 
in $\vert \uparrow_x \rangle$ and $\vert \downarrow_x \rangle$ states and cause two spots on the phosphor screen.
Only $23\%$ of the students provided the correct response. 
The most common difficulty was assuming that because the spin state is $\vert \uparrow_z \rangle$, 
there should not be any splitting as in the following examples: (a) ``Magnetic field gradient cannot affect the electron because it is perpendicular to the wavefunction.'' (b) ``Electrons are undeflected or rather the beam is not split because $\vec B$ is perpendicular to spin state.'' (c) ``The direction of the spin state of the beam of electrons is $y$, and the magnetic field gradient is in the $-x$ direction. The two directions have an angle $90^\circ$, so the magnetic field gradient gives no force to electrons.'' (d) ``With the electrons in only one measurable state, they will experience a force only in one direction upon interaction with $\vec B$.''

Thus, many students explained their reasoning by claiming
that because the magnetic field gradient is in the $-x$ direction but
the spin state is along the $z$ direction, they are orthogonal to each other,
and therefore, there cannot be any splitting of the beam. It is clear from the responses that 
students incorrect relate the direction of the magnetic field in real space 
with the ``direction" of the state vectors in Hilbert space. 
Several students in response to Question~5(b) drew a monotonically increasing function.
One interviewed student drew a diagram of a molecular orbital with four lobes 
and said ``this question asks about the electron cloud pattern due to spin \ldots I am wondering what the spin part of the wavefunction
looks like." Then he added, 
``I am totally blanking on what the plot of $\vert \uparrow_z \rangle$ looks like; otherwise 
I would have done better on this question.'' It is clear from such responses that the abstract nature of spin angular momentum poses special problems in teaching quantum physics.

In comparison to Question~5(a), many more student responses to Question~5(b) mentioned that there would be only one spot on the screen,
but there was no consensus on the direction of the deflection despite the fact that students were asked to ignore the Lorentz force. Some students drew
the spot at the origin, some showed deflections along the positive or negative $x$ direction, and some along the positive or negative $z$ direction. 
They often provided interesting reasons for their choices.
Students were confused about the direction in which the magnetic field gradient would cause the splitting of the beam. The
$13\%$ of the students (including Questions~5(a) and 5(b)) drew the splitting of the beam in the wrong direction (along the $x$ axis
in 5(a) and along the $z$ axis in 5(b)). 
One interviewed student who drew it in the wrong direction said, ``I remember doing this recently and I know there is some splitting 
but I don't remember in which direction it will be."
Another surprising fact is that a large number of students did not respond to Question~5. 
Some explicitly wrote that they don't recall learning about
the Stern-Gerlach experiment. Instructors of the graduate quantum courses 
should take note of the fact that many students may not have done much
related to Stern-Gerlach experiment in the undergraduate quantum mechanics courses.

\subsection{Difficulty in sketching the shape of the Wavefunction}

Questions related to the shape of the wavefunction show that students may not draw a qualitatively correct sketch even if their 
mathematical form of the wavefunction is correct, 
may draw wavefunctions with discontinuities or cusps,
or may confuse a scattering state wavefunction for a potential barrier problem with the wavefunction for a potential well problem.
In Question~2(e) which was related to Question~2(d), students were asked to plot the wavefunction in position space right after the 
measurement of energy in
Question~2(d). We wanted students to draw the wavefunction for the first excited state as a function of position $x$. This wavefunction
is sinusoidal and goes to zero at $x=0, a/2, a$. 
In Question~(6), students were given the potential energy diagram for a finite square well. 
In part (a) they were asked to sketch the ground state wavefunction, and in 
part (b) they had to sketch any one scattering state wavefunction.
In both cases, students were asked to comment on the shape of the wavefunction in the three regions. 
We hoped that in part (a) students would draw the ground
state wavefunction as a sinusoidal curve with no nodes inside the well and with exponentially decaying tails in the classically forbidden regions. 
The wavefunction and its first derivative should be continuous everywhere and the wavefunction should be single valued. In part (b) we expected to see oscillatory behavior in all regions, but because the potential energy is lower in the well, 
the wavelength is shorter in the well. 
For Question~6(b), all responses that were oscillatory in both regions (regardless of the relative wavelengths 
or amplitudes in different regions) and showed the wavefunction and its first derivative as continuous
were considered correct. If the students drew the wavefunction correctly, 
we considered their response correct even if they did not comment
on the shape of the wavefunction in the three regions. 

In Question~2(e), the most common incorrect response, provided by $14\%$ of the students, 
was drawing too many zero crossings in the wavefunction. 
Because many of these students had answered part 2(d) correctly and had written down the wavefunction as $\phi_2(x)$, it is interesting that the mathematical
and graphical representations of $\phi_2(x)$ were inconsistent. 
In response to Question~6(a), $8\%$ of students drew the ground state wavefunction for the infinite square well 
that goes to zero in the classically forbidden
region, and another $8\%$ drew an oscillatory wavefunction in all three regions.
Including both Questions~6(a) and (b), $20\%$ of the students drew either the
first excited state or a higher excited bound state with many oscillations in the well 
and exponential decay outside (a majority of these were in response to Question~6(b)). 
Several students made comments such as ``the particle is bound inside the well but free outside the well.''
The comments displayed confusion about what ``bound state" means and
whether the entire wavefunction is associated with the particle at a
given time or the parts of the wavefunction outside and inside
the well are associated with the particle at different times.
In Question~6(b), approximately $8\%$ of the students 
drew a scattering state wavefunction that had an exponential decay in the well.
Although students were explicitly given a diagram of the potential well, 
they may be confusing the potential well with a potential barrier.
In response to Question~6(a), one interviewed student plotted a wavefunction (without labeling the axes) which looked like 
a parabolic well with the entire function drawn below the horizontal axis. The interviewer then asked whether the wavefunction can have a positive 
amplitude, that is, whether his wavefunction multiplied by an overall minus sign is also a valid ground state wavefunction for this potential well.
The student responded, ``I don't think so. How can the wavefunction not follow the sign of the potential?''
It was apparent from further probing that he was not clear about the fact that the wavefunction
can have an overall complex phase factor.

Including both Questions~6(a) and (b), 
approximately $8\%$ of the students drew wavefunctions with incorrect boundary conditions or that had
discontinuities or cusps in some locations. 
In Question~2(e), $8\%$ of the students had incorrect boundary conditions, for example, 
their wavefunction did not go to zero at $x=0$ and $x=a$ and abruptly ended at 
some non-zero value, or the sinusoidal wavefunction continued beyond $x=0$ and $x=a$ as though it were a free particle for all $x$.

\section{Summary}

The analysis of survey data and interviews indicate that students share common difficulties about quantum mechanics.
All of the questions were qualitative.
They required students to interpret and draw qualitative inferences from quantitative tools
and make a transition from the mathematical representation to concrete cases.
Even if relevant knowledge was not completely lacking, it was often difficult
for students to 
make correct inferences in specific situations.

Shared misconceptions in quantum mechanics can
be traced in large part to incorrect over-generalizations of concepts
learned earlier, compounding of misconceptions that were never cleared up,
or failure to distinguish between closely related concepts. 
Some of the difficulties including those with the time-evolution of the wavefunction
originate from the overemphasis on the time-independent Schroedinger equation and over-generalizing and attributing
the properties of the stationary states to non-stationary states.
Students also had difficulty realizing that all smooth wavefunctions that satisfy the boundary conditions for a system are
possible. Their responses displayed that they do not have a good grasp of Fourier analysis and have difficulty interpreting
that linear superposition of sine functions can result in functions that are not sinusoidal. Students often focused solely on the time-independent Schroedinger equation to
determine if a wavefunction is a possible wavefunction. 
Many students did not know or recall the interpretation of
expectation value as an ensemble average, and did not realize that the
expectation value of an observable can be calculated from the knowledge of the probability of measuring different
values of that observable. 
In the context of the Stern-Gerlach experiment, students had difficulty
distinguishing between vectors in the real space of the laboratory and the state space. 
Many students incorrectly believed that a magnetic field gradient in the $x$ direction cannot
affect a spin-1/2 particle in the eigenstate of $\hat S_z$ because the field gradient and eigenstate of $\hat S_z$
are orthogonal. Students also had difficulty qualitatively sketching the bound state and scattering state wavefunctions
given the potential energy diagram.

Our findings can help the design of better curricula and pedagogies for teaching undergraduate 
quantum mechanics and can inform the instructors of the first-year graduate quantum courses 
of what they can assume about the prior knowledge of the incoming graduate students. 
We are currently developing and assessing quantum interactive learning tutorials 
which incorporate paper and pencil tasks and computer simulations suitable for use in advanced undergraduate quantum mechanics courses 
as supplements to lectures and standard homework assignments.
The goal of the tutorials is to actively engage students in the learning process and help them build links between 
the formalism and the conceptual aspects of quantum physics without compromising the technical content.

\appendix
\section{Survey Questions}

(1) Write down the most fundamental equation of quantum mechanics.\\

In all of the following problems, assume that the measurement of all physical observables is ideal.

(2) The wavefunction of an electron in a one-dimensional infinite square well of width $a$ 
at time $t=0$ 
is given by $\psi(x,0)=\sqrt{2/7} \phi_1(x)+\sqrt{5/7} \phi_2(x)$ where $\phi_1(x)$ and $\phi_2(x)$
are the ground state and first excited stationary state of the system. 
($\phi_n(x)=\sqrt{2/a} \sin(n\pi x/a)$, $E_n=n^2 \pi^2 \hbar^2/(2ma^2)$ where $n=1,2,3 \ldots $)
\begin{center}
\epsfig{file=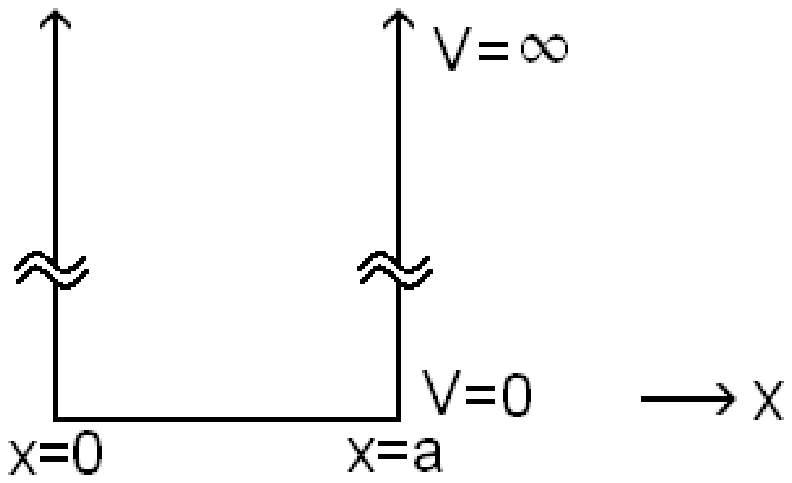,height=1.in}
\end{center}

Answer the following questions about this system:

(a) Write down the wavefunction $\Psi(x,t)$ at time $t$ in terms of $\phi_1(x)$ and $\phi_2(x)$.

(b) You measure the energy of an electron at time $t=0$. Write down the possible values of the energy and the probability 
of measuring each. 

(c) Calculate the expectation value of the energy in the state $\psi(x,t)$ above.

(d) If the energy measurement yields $4\pi^2 \hbar^2/(2ma^2)$, write an expression for the wavefunction right after the measurement.

(e) Sketch the above wavefunction in position space right after the measurement of energy in the previous question.

(f) Immediately after the energy measurement, you measure the position of the electron. Qualitatively describe the possible values of position you can measure and the probability of measuring them.

(3) Which of the following wavefunctions at time $t=0$ are allowed for an electron in a one-dimensional infinite square well of width $a$:
$A\sin^3(\pi x/a)$, $A[\sqrt{2/5} \sin(\pi x/a)+ \sqrt{3/5} \sin(2\pi x/a)]$ and $A e^{-{((x-a/2)/a)}^2}$? In each of the three cases, $A$ is
a suitable normalization constant. You must provide a clear reasoning for each case.
\begin{center}
\epsfig{file=problem3.eps,height=1.in}
\end{center}

(4) Consider the following statement: ``By definition, the Hamiltonian acting on {\it any} allowed state of the system $\psi$ will
give the same state back, that is, $\hat H \psi=E \psi$" where $E$ is the energy of the system. 
Explain why you agree or disagree with this statement.

(5) Notation: $\vert \uparrow_z \rangle$ and $\vert \downarrow_z \rangle$ represent the orthonormal eigenstates of 
$\hat S_z$ (the $z$ component of the spin angular momentum) of the electron. SGA is an abbreviation for a
Stern-Gerlach apparatus. Ignore the Lorentz force on the electron.

(a) A beam of electrons propagating along the $y$ direction (into the page) in spin state $(\vert \uparrow_z \rangle+\vert \downarrow_z \rangle)/\sqrt{2}$
is sent through an SGA with a vertical magnetic field gradient in the $-z$ direction. Sketch the
electron cloud pattern that you expect to see on a distant phosphor screen in the $x$-$z$ plane. Explain your reasoning.
\begin{center}
\epsfig{file=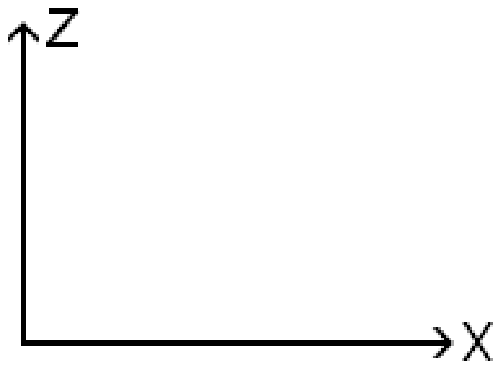,height=1.in}
\end{center}

(b) A beam of electrons propagating along the $y$ direction (into the page) in spin state $\vert \uparrow_z \rangle$
is sent through an SGA with a horizontal magnetic field gradient in the $-x$ direction. 
Sketch the electron cloud pattern that you expect to see on a distant phosphor screen in the $x$-$z$ plane. Explain your reasoning.
\begin{center}
\epsfig{file=problem5a.eps,height=1.in}
\end{center}

(6) The potential energy diagram for a finite square well of width $a$ and depth $-V_0$ is shown below.
\begin{center}
\epsfig{file=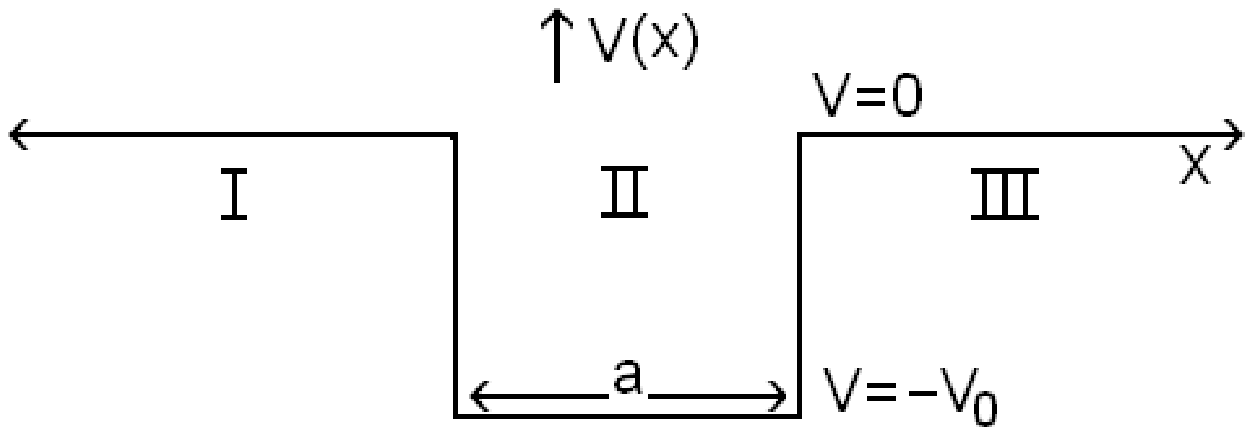,height=1.5in}
\end{center}

(a) Below, draw a qualitative sketch of the ground state wavefunction and comment on the shape of the wavefunction in 
all the three regions shown above.

(b) Draw a qualitative sketch of any one scattering state wavefunction and comment on the shape of the wavefunction in 
all the three regions shown above.

\begin{acknowledgments}

This work is supported in part by the National Science Foundation award PHY-0555434.
We are very grateful to all the faculty and J.\ Gaffney who were consulted during the survey design.
We are also very grateful to all the faculty who administered the survey.
\end{acknowledgments}

\pagebreak

\begin{table}[h]
\centering
\begin{tabular}[t]{|c|c|}
\hline
University&Number of Students\\[0.5 ex]
\hline \hline
The Ohio State University& 49\\[0.5 ex]
\hline
State University of New York (SUNY), Buffalo& 32\\[0.5 ex]
\hline
University of California, Davis& 39\\[0.5 ex]
\hline
University of Iowa& 6\\[0.5 ex]
\hline
University of California, Irvine& 29\\[0.5 ex]
\hline
University of Pittsburgh& 21\\[0.5 ex]
\hline
University of California, Santa Barbara& 26\\[0.5 ex]
\hline
\end{tabular}
\vspace{0.1in}
\caption{\label{1}The number of graduate students from each university that participated in this study.}
\end{table}

\begin{table}[h]
\centering
\begin{tabular}[t]{|c|c|c|}
\hline
Quest$\#$& $\%$Correct&Percentage of Students with Common Difficulties \\[0.5 ex]
\hline \hline
1& 32& (i) 48$\%$ ($\hat H \psi=E \psi$) \\[0.5 ex]
\hline
2(a) &43 &(i) 31$\%$ (common phase factor), \\[0.5 ex]
 & & (ii) 9$\%$ (no time-dependence)\\[0.5 ex]
\hline
2(b)& 67& (i) 7$\%$ (individual measurement versus expectation value)\\[0.5 ex]
\hline
2(c) &39 &(i) 17$\%$ (wrote $\langle \Psi \vert E \vert \Psi \rangle$ or 
 $\langle \Psi \vert H \vert \Psi \rangle$ but nothing else)\\[0.5 ex]
\hline
2(d) &62+17=79 &(i) 17$\%$ (unnormalized wavefunction $\Psi(x)=\sqrt{5/7} \phi_2(x)$) \\[0.5 ex]
\hline
2(e) &56 &(i) 14$\%$ (too many wiggles in wavefunction), \\[0.5 ex]
& & (ii) 8$\%$ (incorrect boundary condition)\\[0.5 ex]
\hline
2 f&38 &(i) 11$\%$ (probability versus expectation value of position $\langle x \rangle$), \\[0.5 ex]
& &(ii) 7$\%$ (incorrect application of uncertainty principle)\\[0.5 ex]
\hline
3&34 &(i) 45$\%$ (first wavefunction not allowed, second allowed), \\[0.5 ex]
& &(ii) 5$\%$ (third wavefunction allowed)\\[0.5 ex]
\hline
4&29 &(i) 39$\%$ (agree with statement unconditionally), \\[0.5 ex]
& &(ii) 11$\%$ (Hamiltonian acting on a state is measurement of energy)\\[0.5 ex]
& &(iii) 10$\%$ (agree with statement if energy is conserved)\\[0.5 ex]
\hline
5(a)& 41& (i) 13$\%$ (splitting into two spots along wrong direction including \\[0.5 ex]
5(b)& 23& questions 5(a) and (b))\\[0.5 ex]
\hline
6(a) &57 &(i) 20$\%$ (first excited or higher excited bound states), \\[0.5 ex]
&&(ii) 8$\%$ (ground state of infinite square well),\\[0.5 ex]
&&(iii) 8$\%$ (incorrect boundary condition), \\[0.5 ex]
&&(iv) 8$\%$ (oscillatory wavefunction in all three regions)\\[0.5 ex]
\hline
6(b)&17 &(i) 8$\%$ (wavefunction with exponential decay inside the well)\\[0.5 ex]
\hline
\end{tabular}
\vspace{0.1in}
\caption{\label{2}Percentage of correct responses on each question and percentage of students with
common difficulties on each question. In the last column,
(i), (ii), (iii) etc. catalog common difficulties for a given question.}
\end{table}

\end{document}